# The Thermal Discrete Dipole Approximation (T-DDA) for near-field radiative heat transfer simulations in three-dimensional arbitrary geometries

Sheila Edalatpour[*] and Mathieu Francoeur[*]

*Radiative Energy Transfer Lab, Department of Mechanical Engineering, University of Utah, Salt Lake City, UT 84112, USA*

**Abstract**

A novel numerical method called the Thermal Discrete Dipole Approximation (T-DDA) is proposed for modeling near-field radiative heat transfer in three-dimensional arbitrary geometries. The T-DDA is conceptually similar to the Discrete Dipole Approximation, except that the incident field originates from thermal oscillations of dipoles. The T-DDA is described in details in the paper, and the method is tested against exact results of radiative conductance between two spheres separated by a sub-wavelength vacuum gap. For all cases considered, the results calculated from the T-DDA are in good agreement with those from the analytical solution. When considering frequency-independent dielectric functions, it is observed that the number of sub-volumes required for convergence increases as the sphere permittivity increases. Additionally, simulations performed for two silica spheres of 0.5 μm-diameter show that the resonant modes are predicted accurately via the T-DDA. For separation gaps of 0.5 μm and 0.2 μm, the relative differences between the T-DDA and the exact results are 0.35% and 6.4%, respectively, when 552 sub-volumes are used to discretize a sphere. Finally, simulations are

---
[*] Corresponding authors. Tel.: +1 801 581 5721, Fax: +1 801 585 9825

E-mail addresses: mfrancoeur@mech.utah.edu (M. Francoeur), sheila.edalatpour@utah.edu (S. Edalatpour)

performed for two cubes of silica separated by a sub-wavelength gap. The results revealed that faster convergence is obtained when considering cubical objects rather than curved geometries. This work suggests that the T-DDA is a robust numerical approach that can be employed for solving a wide variety of near-field thermal radiation problems in three-dimensional geometries.

*Keywords*: Near-field thermal radiation; Thermal Discrete Dipole Approximation (T-DDA); three-dimensional arbitrary geometries; numerical solution of the stochastic Maxwell equations.

# 1. Introduction

Radiation heat transfer between bodies separated by distances greater than the dominant thermal wavelength is limited by Planck's blackbody distribution. In this far-field regime, radiative heat exchange predictions in three-dimensional (3D) complex geometries are tractable using well-established numerical techniques such as the discrete ordinates method and Monte Carlo approaches [1,2]. In the near-field regime of thermal radiation, which refers to the case where bodies are separated by sub-wavelength gaps, heat transfer can exceed by several orders of magnitude the blackbody limit [1,3-14]. The enhancement beyond Planck's distribution is due to the extraneous contribution to energy transport by waves evanescently confined within a distance of about a wavelength normal to the surface of a thermal source. These modes include evanescent waves generated by total internal reflection of a propagating wave at the material-gap interface as well as resonant surface waves such as surface phonon-polaritons and surface plasmon-polaritons.

To account for tunneling of evanescent modes and wave interference, near-field heat transfer problems are modeled using fluctuational electrodynamics [1,3,15]. In this formalism, thermal emission is modeled in Maxwell's equations by stochastic currents that are related to the local temperature of the source via the fluctuation-dissipation theorem. So far, the vast majority of near-field radiative heat transfer predictions have been restricted to simple canonical geometries. This is due to the fact that near-field thermal radiation problems have been mainly solved by deriving analytical expressions for the dyadic Green's functions (DGFs); this approach is referred to as the DGF method. The DGF method provides exact results, but becomes intractable when dealing with 3D arbitrarily-shaped objects. Over the past years, the DGF approach has been applied to various cases: two bulks [4-6,15-19], two films [20-23], two structured surfaces [24], two nanoporous materials [25], one-dimensional layered media [26-28], cylindrical cavity [29], two dipoles [30-32], two large spheres [33,34], dipole-surface [35], dipole-structured surface [36], sphere-surface [34,37], two long cylinders [38], two nanorods [39,40], two gratings [41] and *N* small objects (compared to the wavelength) modeled as electric point dipoles [42].

With the rapid advances in nanofabrication, near-field thermal radiation is becoming an important part of heat transfer engineering. Indeed, near-field thermal radiation may find application in imaging [43], thermophotovoltaic power generation [44-48], nanomanufacturing [49,50], thermal management of electronic devices [51], thermal rectification through a vacuum gap [52,53] and radiative property control [54-56] to name only a few. Due to these numerous potential applications, there is a need for predicting near-field heat exchange in 3D complex geometries. Numerical procedures, namely the finite-difference time-domain (FDTD) method [57-59], the finite-difference frequency-domain (FDFD) method [60] and the boundary element method (BEM) [61] have been applied recently to near-field thermal radiation calculations. Both FDTD and FDFD approaches suffer from large computational time, while the BEM is difficult to apply when dealing with heterogeneous materials. In this work, the Thermal Discrete Dipole Approximation (T-DDA) is proposed for simulating near-field heat transfer between 3D arbitrarily-shaped objects. The Discrete Dipole Approximation (DDA), extensively used for predicting electromagnetic wave scattering by particles, is based on discretizing objects into cubical sub-volumes behaving as electric point dipoles [62-66]. The T-DDA follows the same general procedure as the DDA, except that the incident field is induced by thermal fluctuations of dipoles instead of being produced by an external illumination.

The objective of this paper is to formulate the T-DDA and to test the method against exact results obtained from the DGF approach. In the next section, the physical and mathematical formulation of the problem is provided. Next, the T-DDA is derived starting from the stochastic Maxwell equations and the associated solution procedure is detailed. In the fourth section, the T-DDA is verified against exact results for two spheres separated by a sub-wavelength vacuum gap; a problem involving two cubes is presented afterwards. Concluding remarks are finally provided.

**2. Physical and mathematical formulation of the problem**

The problem under consideration is shown schematically in Fig. 1. A total of $m = 1, 2,…, M$ objects at temperatures $T_m$ are submerged in vacuum (medium 0) and are exchanging thermal radiation. The $M_e$ emitters are made up of source points $\mathbf{r}'$ while the $M_a$ absorbers are composed of points $\mathbf{r}$ where the fields are calculated. The bodies are assumed to be in local thermodynamic equilibrium, isotropic, non-magnetic and their electromagnetic responses are described by frequency-dependent dielectric functions $\varepsilon_m = \varepsilon'_m + i\varepsilon''_m$ local in space. No assumptions are made on the shape and size of the objects as well as on their separation distances.

[Insert Fig. 1 about here]

Thermal emission is the result of random fluctuations of charged particles inside a body caused by thermal agitation [3]. On a macroscopic level, this chaotic motion of charged particles is modeled via a stochastic current density $\mathbf{J}^r$ which is added in Maxwell's equations to model thermal emission. The random nature of the thermal current thus makes the Maxwell equations stochastic. Assuming $e^{-i\omega t}$ for the time harmonic fields, the stochastic Maxwell equations in the frequency domain are written as follows [67]:

$$\nabla \times \mathbf{E}(\mathbf{r},\omega) = i\omega\mu_0 \mathbf{H}(\mathbf{r},\omega) \tag{2.1a}$$

$$\nabla \times \mathbf{H}(\mathbf{r},\omega) = -i\omega\varepsilon_0\varepsilon\mathbf{E}(\mathbf{r},\omega) + \mathbf{J}^r(\mathbf{r},\omega) \tag{2.1b}$$

where $\varepsilon_0$ and $\mu_0$ are the vacuum permittivity and permeability, respectively, $\mathbf{E}$ is the electric field and $\mathbf{H}$ denotes the magnetic field.

The first moment of the thermal current is zero (i.e., $\langle \mathbf{J}^r(\mathbf{r},\omega) \rangle = 0$), which implies that the mean radiated fields are also zero [3]. In heat transfer analysis, the quantities of interest are not the mean radiated fields, but the flux and the energy density which are proportional to the ensemble average of the spatial correlation function of currents. This correlation function is provided by the fluctuation-dissipation theorem linking the thermal current to the local temperature of the emitter [3]:

$$\langle \mathbf{J}^r(\mathbf{r}',\omega) \otimes \mathbf{J}^r(\mathbf{r}'',\omega) \rangle = \frac{4\omega\varepsilon_0\varepsilon''}{\pi} \Theta(\omega,T)\delta(\mathbf{r}'-\mathbf{r}'')\bar{\bar{\mathbf{I}}} \tag{2.2}$$

The symbol $\otimes$ represents the outer product defined as $\mathbf{J}^r(\mathbf{r}',\omega) \otimes \mathbf{J}^r(\mathbf{r}'',\omega) = \mathbf{J}^r(\mathbf{r}',\omega) \cdot (\mathbf{J}^r(\mathbf{r}'',\omega))^\dagger$, where the superscript † indicates the Hermitian operator (conjugate transpose). In Eq. (2.2), $\bar{\bar{\mathbf{I}}}$ is the unit dyadic and $\Theta$ is the mean energy of an electromagnetic state given by [68]:

$$\Theta(\omega,T) = \frac{\hbar\omega}{\exp(\hbar\omega/k_B T) - 1} \tag{2.3}$$

where $\hbar$ is the reduced Planck's constant ($= 1.0546 \times 10^{-34}$ J·s) while $k_B$ ($= 1.3807 \times 10^{-23}$ J·K$^{-1}$) is the Boltzmann constant. Note that a factor four is included in the fluctuation-dissipation theorem to account for the fact that only positive frequencies are considered when passing from the time to the frequency domain [68].

In the next section, a general formulation is proposed for modeling near-field radiative heat transfer via the T-DDA.

**3. Description of the T-DDA**

*3.1. Volume integral equation for radiation problems*

The starting point of the T-DDA formulation is to decompose the (total) electric field **E** as the sum of an incident field, **E**$_{inc}$, and a scattered field, **E**$_{sca}$:

$$\mathbf{E}(\mathbf{r},\omega) = \mathbf{E}_{inc}(\mathbf{r},\omega) + \mathbf{E}_{sca}(\mathbf{r},\omega) \tag{3.1}$$

The incident field can be interpreted as the field thermally generated by point sources that is propagating in free space in the absence of scatterers. A volume integral equation for the electric field **E** can be determined by first taking the curl of both sides of Eq. (2.1a):

$$\nabla \times \mathbf{H}(\mathbf{r},\omega) = -\frac{i}{\omega\mu_0} \nabla \times \nabla \times \mathbf{E}(\mathbf{r},\omega) \tag{3.2}$$

The vector wave equation is determined by substituting Eq. (3.2) into Eq. (2.1b):

$$\nabla \times \nabla \times \mathbf{E}(\mathbf{r},\omega) - k^2 \mathbf{E}(\mathbf{r},\omega) = i\omega\mu_0 \mathbf{J}^r(\mathbf{r},\omega) \tag{3.3}$$

where $k$ ($= \omega\sqrt{\varepsilon\varepsilon_0\mu_0}$) is the magnitude of the wavevector. A free space form of the vector wave equation is obtained by subtracting $k_0^2 \mathbf{E}$ from both sides of Eq. (3.3):

$$\nabla \times \nabla \times \mathbf{E}(\mathbf{r},\omega) - k_0^2 \mathbf{E}(\mathbf{r},\omega) = (k^2 - k_0^2)\mathbf{E}(\mathbf{r},\omega) + i\omega\mu_0 \mathbf{J}^r(\mathbf{r},\omega) \tag{3.4}$$

where $k_0$ is the magnitude of the wavevector in vacuum. In typical DDA formulations where particles are illuminated by an external source, the incident field satisfies $\nabla \times \nabla \times \mathbf{E}_{inc} - k_0^2 \mathbf{E}_{inc} = 0$. In radiation heat transfer, the situation is different as the incident field is generated by thermally fluctuating currents. As such, the thermally generated incident field propagating in free space satisfies the following vector wave equation:

$$\nabla \times \nabla \times \mathbf{E}_{inc}(\mathbf{r},\omega) - k_0^2 \mathbf{E}_{inc}(\mathbf{r},\omega) = i\omega\mu_0 \mathbf{J}^r(\mathbf{r},\omega) \tag{3.5}$$

The vector wave equation for the scattered field can then be obtained by subtracting Eq. (3.5) from Eq. (3.4):

$$\nabla \times \nabla \times \mathbf{E}_{sca}(\mathbf{r},\omega) - k_0^2 \mathbf{E}_{sca}(\mathbf{r},\omega) = (k^2 - k_0^2)\mathbf{E}(\mathbf{r},\omega) \tag{3.6}$$

The scattered field in Eq. (3.6) can be interpreted as the field generated by an equivalent source function, $(k^2 - k_0^2)\mathbf{E}$, propagating in free space. Solutions for Eqs. (3.5) and (3.6) are obtained using the free space DGF denoted by $\overline{\overline{\mathbf{G}}}$ [64]:

$$\mathbf{E}_{inc}(\mathbf{r},\omega) = i\omega\mu_0 \int_{V_e} \overline{\overline{\mathbf{G}}}(\mathbf{r},\mathbf{r}',\omega) \cdot \mathbf{J}^r(\mathbf{r}',\omega) dV' \tag{3.7a}$$

$$\mathbf{E}_{sca}(\mathbf{r},\omega) = \int_V (k^2 - k_0^2) \overline{\overline{\mathbf{G}}}(\mathbf{r},\mathbf{r}',\omega) \cdot \mathbf{E}(\mathbf{r}',\omega) dV' \tag{3.7b}$$

where $V_e$ is the volume of the emitting bodies, while $V$ (= $V_e$ + $V_a$) is the total volume where $V_a$ is the volume of the absorbing bodies. It can be seen in Eq. (3.7a) that the integration is performed over $V_e$ only where the thermal source is non-zero, while the integration in Eq. (3.7b) is performed over the total volume $V$ to account for the interactions between all objects. The free space DGF is given by [69]:

$$\overline{\overline{\mathbf{G}}}(\mathbf{r},\mathbf{r}',\omega) = \frac{e^{ik_0 R}}{4\pi R}\left[\left(1 - \frac{1}{(k_0 R)^2} + \frac{i}{k_0 R}\right)\overline{\overline{\mathbf{I}}} - \left(1 - \frac{3}{(k_0 R)^2} + \frac{3i}{k_0 R}\right)\hat{\mathbf{R}} \otimes \hat{\mathbf{R}}\right] \tag{3.8}$$

where $R = |\mathbf{r} - \mathbf{r}'|$ and $\hat{\mathbf{R}} = (\mathbf{r} - \mathbf{r}')/|\mathbf{r} - \mathbf{r}'|$.

A volume integral equation for the electric field is obtained by substituting the scattered field given by Eq. (3.7b) into Eq. (3.1):

$$\mathbf{E}(\mathbf{r},\omega) - k_0^2 \int_V [\varepsilon(\mathbf{r}') - 1] \overline{\overline{\mathbf{G}}}(\mathbf{r},\mathbf{r}',\omega) \cdot \mathbf{E}(\mathbf{r}',\omega) dV' = \mathbf{E}_{inc}(\mathbf{r},\omega) \qquad (3.9)$$

The DGF has a singularity at $\mathbf{r} = \mathbf{r}'$, such that the principal value method is used to circumvent this problem. In this approach, an infinitesimal volume containing the singularity point is excluded from the integral. For a spherical or cubic exclusion volume, the application of the principal value method leads to [64,70]:

$$\begin{aligned} &k_0^2 \int_V [\varepsilon(\mathbf{r}') - 1] \overline{\overline{\mathbf{G}}}(\mathbf{r},\mathbf{r}',\omega) \cdot \mathbf{E}(\mathbf{r}',\omega) dV' \\ &= k_0^2 P.V. \int_V [\varepsilon(\mathbf{r}') - 1] \overline{\overline{\mathbf{G}}}(\mathbf{r},\mathbf{r}',\omega) \cdot \mathbf{E}(\mathbf{r}',\omega) dV' - \frac{\varepsilon(\mathbf{r}) - 1}{3} \mathbf{E}(\mathbf{r},\omega) \end{aligned} \qquad (3.10)$$

where *P.V.* stands for principal value. The core equation underlying the T-DDA method is finally obtained by substituting Eq. (3.10) into Eq. (3.9):

$$\frac{\varepsilon(\mathbf{r}) + 2}{3} \mathbf{E}(\mathbf{r},\omega) - k_0^2 P.V. \int_V [\varepsilon(\mathbf{r}') - 1] \overline{\overline{\mathbf{G}}}(\mathbf{r},\mathbf{r}',\omega) \cdot \mathbf{E}(\mathbf{r}',\omega) dV' = \mathbf{E}_{inc}(\mathbf{r},\omega) \qquad (3.11)$$

where the incident electric field $\mathbf{E}_{inc}$ is specified by Eq. (3.7a). In the next section, Eq. (3.11) is discretized in order to derive a system of linear equations.

*3.2. Discretization of the volume integral equation*

The first step toward the numerical solution of Eq. (3.11) is the discretization of the emitting and absorbing objects into $N$ cubical sub-volumes on a cubical lattice. The first $N_e$ sub-volumes are located in the emitters, while the $N_a$ (= $N$ - $N_e$) remaining sub-volumes are allocated to the absorbers. The discretization should be fine enough such that the dimension of each sub-volume is small compared to the radiation wavelength (more details about the discretization are provided in section 4) [63]. If this condition is satisfied, it can be assumed that the electromagnetic properties and the electric field are uniform inside each sub-volume. Equation (3.11) evaluated at the center $\mathbf{r}_i$ of a sub-volume $i$ can therefore be written as:

$$\frac{\varepsilon_i+2}{3}\mathbf{E}_i - k_0^2 \sum_{j=1}^{N}(\varepsilon_j-1)\left(P.V.\int_{\Delta V_j}\overline{\overline{\mathbf{G}}}(\mathbf{r}_i,\mathbf{r}',\omega)dV'\right)\cdot\mathbf{E}_j = \mathbf{E}_{inc,i}, i = 1, 2, ..., N \tag{3.12}$$

where the subscripts $i$ and $j$ refer to sub-volumes. In Eq. (3.12), when $i \neq j$, the DGF has no singularity, such that the principal value can be approximated as [64]:

$$P.V.\int_{\Delta V_j}\overline{\overline{\mathbf{G}}}(\mathbf{r}_i,\mathbf{r}',\omega)dV' = \overline{\overline{\mathbf{G}}}_{ij}\Delta V_j, i \neq j \tag{3.13}$$

where $\overline{\overline{\mathbf{G}}}_{ij}$ is an abbreviation for $\overline{\overline{\mathbf{G}}}(\mathbf{r}_i,\mathbf{r}_j)$. For $i = j$, the principal value integral is given by [64]:

$$P.V.\int_{\Delta V_j}\overline{\overline{\mathbf{G}}}(\mathbf{r}_i,\mathbf{r}',\omega)dV' = \frac{2}{3k_0^2}\left[e^{ik_0 a_i}(1-ik_0 a_i)-1\right]\overline{\overline{\mathbf{I}}}, i = j \tag{3.14}$$

where $a_i$ is the effective radius of sub-volume $i$ defined as $(3\Delta V_i/4\pi)^{1/3}$. The discretized version of the volume integral equation is obtained by substituting Eqs. (3.13) and (3.14) into Eq. (3.12):

$$\left[\frac{\varepsilon_i+2}{3}-\frac{2(\varepsilon_i-1)}{3}\left(e^{ik_0 a_i}(1-ik_0 a_i)-1\right)\right]\mathbf{E}_i - k_0^2\sum_{\substack{j=1\\j\neq i}}^{N}(\varepsilon_j-1)\Delta V_j\overline{\overline{\mathbf{G}}}_{ij}\cdot\mathbf{E}_j = \mathbf{E}_{inc}, i = 1, 2, ..., N \tag{3.15}$$

Equation (3.15) is a system of $N$ vector equations where the electric field in each sub-volume is the unknown. It is important to keep in mind that $\mathbf{E}_i$ is stochastic since the incident field is generated by random currents. The incident field, given by Eq. (3.7a), is approximated as follows after discretization into sub-volumes:

$$\mathbf{E}_{inc,i} = \begin{cases} 0 & i = 1, 2, ..., N_e \\ i\omega\mu_0 \sum_{k=1}^{N_e} \overline{\overline{\mathbf{G}}}_{ik} \cdot \mathbf{J}_k^r \Delta V_k & i = N_e + 1, N_e + 2, ..., N \end{cases} \quad (3.16)$$

Equation (3.16) stipulates that the thermally generated incident field is nil in the emitters ($i = 1, 2, ..., N_e$), while the incident field within the absorbers ($i = N_e + 1, N_e + 2, ..., N$) is due to the $k = 1, 2, ..., N_e$ emitting dipoles.

From now on, it is assumed that each sub-volume is behaving as an electric point dipole, such that Eq. (3.15) can be re-written in terms of unknown equivalent dipole moments $\mathbf{p}_i$ instead of unknown electric fields $\mathbf{E}_i$ using the relation [64]:

$$\mathbf{E}_i = \frac{3}{\alpha_i^{CM}(\varepsilon_i + 2)} \mathbf{p}_i \quad (3.17)$$

where $\alpha_i^{CM}$ is the Clausius-Mossotti polarizability defined as:

$$\alpha_i^{CM} = 3\varepsilon_0 \frac{\varepsilon_i - 1}{\varepsilon_i + 2} \Delta V_i \quad (3.18)$$

The random current density $\mathbf{J}_k^r$ in Eq. (3.16) can also be expressed in terms of a thermally fluctuating dipole moment $\mathbf{p}_k^r$ representing thermal emission [69]:

$$\mathbf{J}_k^r = \frac{-i\omega}{\Delta V_k} \mathbf{p}_k^r \quad (3.19)$$

As for the random current, the mean of the thermally fluctuating dipole moment is equal to zero. The random dipole moments are related to the local temperature of the medium via a modified version of the fluctuation-dissipation theorem (see Eq. (2.2)) [71]:

$$\langle \mathbf{p}_k^r \otimes \mathbf{p}_k^r \rangle = \frac{4\varepsilon_0 \operatorname{Im}(\alpha_k^{CM})}{\pi\omega} \Theta(\omega, T) \bar{\bar{\mathbf{I}}} \tag{3.20}$$

Using Eqs. (3.17) through (3.19), the system of equations given by Eq. (3.15) can be re-written in terms of unknown dipole moments $\mathbf{p}_i$:

$$\frac{1}{\alpha_i} \mathbf{p}_i - \frac{k_0^2}{\varepsilon_0} \sum_{\substack{j=1 \\ j \neq i}}^{N} \bar{\bar{\mathbf{G}}}_{ij} \cdot \mathbf{p}_j = \mathbf{E}_{inc,i}, \quad i = 1,2,\ldots,N \tag{3.21}$$

where

$$\mathbf{E}_{inc,i} = \begin{cases} 0 & i = 1,2,\ldots,N_e \\ \mu_0 \omega^2 \sum_{k=1}^{N_e} \bar{\bar{\mathbf{G}}}_{ik} \cdot \mathbf{p}_k^r & i = N_e+1, N_e+2,\ldots,N \end{cases} \tag{3.22}$$

The variable $\alpha_i$ in Eq. (3.21) is referred to as the radiative polarizability of dipole $i$, and is defined as [64]:

$$\alpha_i = \frac{\alpha_i^{CM}}{1 - \frac{\alpha_i^{CM}}{2\pi\varepsilon_0 a_i^3}\left[e^{ik_0 a_i}(1 - ik_0 a_i) - 1\right]} \tag{3.23}$$

Equation (3.21) can be interpreted as follows. The first term on the left-hand side represents the interaction of dipole $i$ with itself (i.e., self-interaction term), while the second term accounts for the interactions of dipole $i$ with all other dipoles except $j = i$. The right-hand side of Eq. (3.21) is the incident field in the absorbing dipoles ($i = N_e + 1, N_e + 2, \ldots, N$) due to thermal emission by the emitting dipoles ($i = 1, 2, \ldots, N_e$). The system of $3N$ scalar equations with $3N$ unknowns (each dipole $i$ has three orthogonal components) can also be written in a compact form using the following matrix notation:

$$\overline{\overline{\mathbf{A}}} \cdot \overline{\mathbf{P}} = \overline{\mathbf{E}}_{inc} \tag{3.24}$$

where $\overline{\mathbf{P}}$ is the 3$N$ stochastic column vector containing the unknown dipole moments $\mathbf{p}_i$, $\overline{\mathbf{E}}_{inc}$ is the 3$N$ stochastic column vector containing the known incident fields $\mathbf{E}_{inc,i}$, while $\overline{\overline{\mathbf{A}}}$ is the 3$N$ by 3$N$ deterministic interaction matrix consisting of $N^2$ 3 by 3 sub-matrices. Each sub-matrix $\overline{\overline{\mathbf{A}}}_{ij}$ represents the interactions between dipoles $i$ and $j$. For clarity, the expanded form of Eq. (3.24) is also given:

$$\begin{bmatrix} \overline{\overline{\mathbf{A}}}_{11} & \overline{\overline{\mathbf{A}}}_{12} & \cdots & \overline{\overline{\mathbf{A}}}_{1N} \\ \overline{\overline{\mathbf{A}}}_{21} & \overline{\overline{\mathbf{A}}}_{22} & \cdots & \overline{\overline{\mathbf{A}}}_{2N} \\ \vdots & \vdots & \ddots & \vdots \\ \overline{\overline{\mathbf{A}}}_{N1} & \overline{\overline{\mathbf{A}}}_{N2} & \cdots & \overline{\overline{\mathbf{A}}}_{NN} \end{bmatrix} \begin{bmatrix} \mathbf{p}_1 \\ \mathbf{p}_2 \\ \vdots \\ \mathbf{p}_N \end{bmatrix} = \begin{bmatrix} \mathbf{E}_{inc,1} \\ \mathbf{E}_{inc,2} \\ \vdots \\ \mathbf{E}_{inc,N} \end{bmatrix} \tag{3.25}$$

For $i \neq j$, the sub-matrix $\overline{\overline{\mathbf{A}}}_{ij}$ is derived from Eq. (3.21) combined with Eq. (3.8):

$$\overline{\overline{\mathbf{A}}}_{ij} = C_{ij} \begin{bmatrix} \beta_{ij} + \gamma_{ij} \hat{r}_{ij,x}^2 & \gamma_{ij} \hat{r}_{ij,x} \hat{r}_{ij,y} & \gamma_{ij} \hat{r}_{ij,x} \hat{r}_{ij,z} \\ \gamma_{ij} \hat{r}_{ij,y} \hat{r}_{ij,x} & \beta_{ij} + \gamma_{ij} \hat{r}_{ij,y}^2 & \gamma_{ij} \hat{r}_{ij,y} \hat{r}_{ij,z} \\ \gamma_{ij} \hat{r}_{ij,z} \hat{r}_{ij,x} & \gamma_{ij} \hat{r}_{ij,z} \hat{r}_{ij,y} & \beta_{ij} + \gamma_{ij} \hat{r}_{ij,z}^2 \end{bmatrix}, \; i \neq j \tag{3.26}$$

where

$$\hat{r}_{ij,\alpha} = \frac{r_{ij,\alpha}}{r_{ij}}, \; \alpha = x, y, z \tag{3.27a}$$

$$C_{ij} = -\frac{k_0^2}{4\pi\varepsilon_0} \frac{e^{ik_0 r_{ij}}}{r_{ij}} \tag{3.27b}$$

$$\beta_{ij} = \left[ 1 - \frac{1}{(k_0 r_{ij})^2} + \frac{i}{k_0 r_{ij}} \right] \tag{3.27c}$$

$$\gamma_{ij} = -\left[ 1 - \frac{3}{(k_0 r_{ij})^2} + \frac{3i}{k_0 r_{ij}} \right] \tag{3.27d}$$

Note that $r_{ij}$ is the magnitude of the distance vector $\mathbf{r}_{ij}$ between dipoles $i$ and $j$, while $\hat{\mathbf{r}}_{ij}$ is the unit vector along $\mathbf{r}_{ij}$.

For the self-interaction term ($i = j$), the sub-matrix $\overline{\overline{\mathbf{A}}}_{ii}$ is given by:

$$\overline{\overline{\mathbf{A}}}_{ii} = \frac{1}{\alpha_i} \overline{\overline{\mathbf{I}}} \tag{3.28}$$

In the next section, the solution of the stochastic system of equations is discussed.

*3.3. Heat transfer calculations*

The main objective in heat transfer calculations is to compute the radiative power exchanged between objects. The mean energy dissipated in the absorbers ($i = N_e + 1, N_e + 2, \ldots, N$) at a given frequency is calculated as [63,64]:

$$\langle Q_{abs,\omega} \rangle = \frac{\omega}{2} \sum_{i=N_e+1}^{N} \left( \text{Im}[(\alpha_i^{-1})^*] - \frac{2}{3} k_0^2 \right) \text{tr}(\langle \mathbf{p}_i \otimes \mathbf{p}_i \rangle) \tag{3.29}$$

According to Eq. (3.29), the unknown dipole moments $\mathbf{p}_i$ do not need to be calculated directly. Instead, the trace of the dipole auto-correlation function, $\text{tr}(\langle \mathbf{p}_i \otimes \mathbf{p}_i \rangle)$, is needed in order to compute the power absorbed. The procedure for calculating the auto-correlations is described hereafter.

If the interaction matrix $\overline{\overline{\mathbf{A}}}$ is invertible, the dipole moment vector $\overline{\mathbf{P}}$ can be determined from Eq. (3.24):

$$\overline{\mathbf{P}} = \overline{\overline{\mathbf{A}}}^{-1} \cdot \overline{\mathbf{E}}_{inc} \tag{3.30}$$

where $\overline{\overline{\mathbf{A}}}^{-1}$ is the inverse of matrix $\overline{\overline{\mathbf{A}}}$. The mean value of $\overline{\mathbf{P}}$ can then be determined from Eq. (3.30):

$$\langle \overline{\mathbf{P}} \rangle = \overline{\overline{\mathbf{A}}}^{-1} \cdot \langle \overline{\mathbf{E}}_{inc} \rangle \tag{3.31}$$

where $\overline{\overline{\mathbf{A}}}^{-1}$ is taken out of the mean operator since it is a deterministic matrix. Using Eq. (3.22), the mean of the incident field at a given dipole $i$ is:

$$\langle \mathbf{E}_{inc,i} \rangle = \begin{cases} 0 & i = 1, 2, .., N_e \\ \mu_0 \omega^2 \sum_{k=1}^{N_e} \overline{\overline{\mathbf{G}}}_{ik} \cdot \langle \mathbf{p}_k^r \rangle = 0 & i = N_e + 1, N_e + 2, ..., N \end{cases} \tag{3.32}$$

where the linear property of mean operator has been utilized. Equation (3.32) shows that the mean incident field is zero regardless of $i$, since the first moment of the thermally fluctuating dipoles is zero. As a result, the mean of the dipole moment vector, $\langle \overline{\mathbf{P}} \rangle$, is equal to zero.

Equation (3.30) is also used to calculate the correlation matrix of $\overline{\mathbf{P}}$. The correlation matrix of the zero-mean dipole moment vector is defined as [72,73]:

$$\overline{\overline{\mathbf{R}}}_{\mathbf{PP}} = \langle \overline{\mathbf{P}} \otimes \overline{\mathbf{P}} \rangle \tag{3.33}$$

where $\overline{\overline{\mathbf{R}}}_{\mathbf{PP}}$ is a $3N$ by $3N$ matrix consisting of $N^2$ sub-matrices:

$$\overline{\overline{\mathbf{R}}}_{\mathbf{PP}} = \begin{bmatrix} \overline{\overline{\mathbf{R}}}_{\mathbf{p}_1 \mathbf{p}_1} & \overline{\overline{\mathbf{R}}}_{\mathbf{p}_1 \mathbf{p}_2} & \cdots & \overline{\overline{\mathbf{R}}}_{\mathbf{p}_1 \mathbf{p}_N} \\ \overline{\overline{\mathbf{R}}}_{\mathbf{p}_2 \mathbf{p}_1} & \overline{\overline{\mathbf{R}}}_{\mathbf{p}_2 \mathbf{p}_2} & \cdots & \overline{\overline{\mathbf{R}}}_{\mathbf{p}_2 \mathbf{p}_N} \\ \vdots & \vdots & \ddots & \vdots \\ \overline{\overline{\mathbf{R}}}_{\mathbf{p}_N \mathbf{p}_1} & \overline{\overline{\mathbf{R}}}_{\mathbf{p}_N \mathbf{p}_2} & \cdots & \overline{\overline{\mathbf{R}}}_{\mathbf{p}_N \mathbf{p}_N} \end{bmatrix} \tag{3.34}$$

A given 3 by 3 sub-matrix, $\bar{\bar{\mathbf{R}}}_{\mathbf{p}_i\mathbf{p}_j} = \langle \mathbf{p}_i \otimes \mathbf{p}_j \rangle$, is the correlation matrix of the dipole moments $\mathbf{p}_i$ and $\mathbf{p}_j$. For calculating the power absorbed, only the traces of the correlation matrices $\bar{\bar{\mathbf{R}}}_{\mathbf{p}_i\mathbf{p}_i}$ are needed, such that Eq. (3.29) can be re-written as:

$$\langle Q_{abs,\omega} \rangle = \frac{\omega}{2} \sum_{i=N_e+1}^{N} \left( \mathrm{Im}[(\alpha_i^{-1})^*] - \frac{2}{3}k_0^2 \right) \mathrm{tr}\left(\bar{\bar{\mathbf{R}}}_{\mathbf{p}_i\mathbf{p}_i}\right) \tag{3.35}$$

Substitution of Eq. (3.30) into Eq. (3.33), and using the identity $(\bar{\bar{\mathbf{A}}}^{-1} \cdot \mathbf{E}_{inc})^\dagger = \mathbf{E}_{inc}^\dagger \cdot (\bar{\bar{\mathbf{A}}}^{-1})^\dagger$, the correlation matrix of $\bar{\mathbf{P}}$ can be written as:

$$\bar{\bar{\mathbf{R}}}_{\mathbf{PP}} = \bar{\bar{\mathbf{A}}}^{-1} \cdot \bar{\bar{\mathbf{R}}}_{\mathbf{EE}} \cdot \left(\bar{\bar{\mathbf{A}}}^{-1}\right)^\dagger \tag{3.36}$$

where $\bar{\bar{\mathbf{R}}}_{\mathbf{EE}}$ is the $3N$ by $3N$ correlation matrix of the incident field consisting of $N^2$ sub-matrices. Using the fact that the mean value of the incident field is zero, $\bar{\bar{\mathbf{R}}}_{\mathbf{EE}}$ is calculated as:

$$\bar{\bar{\mathbf{R}}}_{\mathbf{EE}} = \langle \bar{\mathbf{E}}_{inc} \otimes \bar{\mathbf{E}}_{inc} \rangle = \begin{bmatrix} \bar{\bar{\mathbf{R}}}_{\mathbf{E}_1\mathbf{E}_1} & \bar{\bar{\mathbf{R}}}_{\mathbf{E}_1\mathbf{E}_2} & \cdots & \bar{\bar{\mathbf{R}}}_{\mathbf{E}_1\mathbf{E}_N} \\ \bar{\bar{\mathbf{R}}}_{\mathbf{E}_2\mathbf{E}_{11}} & \bar{\bar{\mathbf{R}}}_{\mathbf{E}_2\mathbf{E}_2} & \cdots & \bar{\bar{\mathbf{R}}}_{\mathbf{E}_2\mathbf{E}_N} \\ \vdots & \vdots & \ddots & \vdots \\ \bar{\bar{\mathbf{R}}}_{\mathbf{E}_N\mathbf{E}_1} & \bar{\bar{\mathbf{R}}}_{\mathbf{E}_N\mathbf{E}_2} & \cdots & \bar{\bar{\mathbf{R}}}_{\mathbf{E}_N\mathbf{E}_N} \end{bmatrix} \tag{3.37}$$

where the 3 by 3 sub-matrix, $\bar{\bar{\mathbf{R}}}_{\mathbf{E}_i\mathbf{E}_j} = \langle \mathbf{E}_{inc,i} \otimes \mathbf{E}_{inc,j} \rangle$, is the correlation matrix of the incident fields in dipoles $i$ and $j$. A given sub-matrix is calculated by substituting $\mathbf{E}_{inc,i}$ and $\mathbf{E}_{inc,j}$ given by Eq. (3.22):

$$\bar{\bar{\mathbf{R}}}_{\mathbf{E}_i\mathbf{E}_j} = \mu_0^2 \omega^4 \sum_{k=1}^{N_e} \sum_{n=1}^{N_e} \bar{\bar{\mathbf{G}}}_{ik} \cdot \langle \mathbf{p}_k^r \otimes \mathbf{p}_n^r \rangle \cdot \bar{\bar{\mathbf{G}}}_{jn}^\dagger, \; i,j \geq N_e + 1 \tag{3.38}$$

where $\langle \mathbf{p}_k^r \otimes \mathbf{p}_n^r \rangle$ is non-zero only for $n = k$, and is given by the fluctuation-dissipation theorem (Eq. (3.20)). Substitution of Eq. (3.20) into Eq. (3.38) leads to the following correlation matrix for the incident field:

$$\bar{\bar{\mathbf{R}}}_{\mathbf{E}_i \mathbf{E}_j} = \frac{4\varepsilon_0 \mu_0^2 \omega^3 \Theta(\omega, T)}{\pi} \sum_{k=1}^{N_e} \text{Im}(\alpha_k^{CM}) \bar{\bar{\mathbf{G}}}_{ik} \cdot \bar{\bar{\mathbf{G}}}_{jk}^\dagger, \, i, j \geq N_e + 1 \qquad (3.39)$$

The DGFs in Eq. (3.39) have already been calculated when determining the interaction matrix (see Eqs. (3.21) and (3.25)). As such, there is no need to re-compute these DGFs. The DGF $\bar{\bar{\mathbf{G}}}_{lk}$ for $l \neq k$ ($l = i, j$) is related to the interaction sub-matrix $\bar{\bar{\mathbf{A}}}_{lk}$ as follows:

$$\bar{\bar{\mathbf{G}}}_{lk} = -\frac{\varepsilon_0}{k_0^2} \bar{\bar{\mathbf{A}}}_{lk}, \, l \neq k \qquad (3.40)$$

The incident field correlation sub-matrix is thus expressed in terms of interaction sub-matrices instead of DGFs by substituting Eq. (3.40) into Eq. (3.39):

$$\bar{\bar{\mathbf{R}}}_{\mathbf{E}_i \mathbf{E}_j} = \frac{4\varepsilon_0 \Theta(\omega, T)}{\pi \omega} \sum_{k=1}^{N_e} \text{Im}(\alpha_k^{CM}) \bar{\bar{\mathbf{A}}}_{ik} \cdot \bar{\bar{\mathbf{A}}}_{jk}^\dagger, \, i, j \geq N_e + 1 \qquad (3.41)$$

Equation (3.41) is employed to populate the correlation matrix given by Eq. (3.37). The correlation matrix $\bar{\bar{\mathbf{R}}}_{\mathbf{EE}}$ is in turn substituted into Eq. (3.36) in order to compute $\bar{\bar{\mathbf{R}}}_{\mathbf{PP}}$. The diagonal elements of $\bar{\bar{\mathbf{R}}}_{\mathbf{PP}}$, corresponding to the auto-correlation of the dipole moments, are finally used for calculating the power absorbed by the $M_a$ objects via Eq. (3.35). From this result, the power absorbed by the $M_e$ objects caused by thermal emission from the $M_a$ bodies can easily be obtained due to the reciprocity of the DGF. As such, the net radiative heat transfer between the emitters and the absorbers can be calculated.

The description of the T-DDA is completed. For clarity, the algorithm for near-field radiation heat transfer predictions via the T-DDA is summarized below.

1. Discretize the emitters and the absorbers into $N_e$ and $N_a$ cubical sub-volumes $\Delta V_i$ conceptualized as electric point dipoles with effective radius $a_i$.

2. Assign a dielectric function $\varepsilon_i$ and a temperature $T_i$ to each sub-volume.

3. Calculate the polarizability $\alpha_i$ of each sub-volume using Eqs. (3.18) and (3.23).

4. Calculate the interaction matrix $\bar{\bar{\mathbf{A}}}$ using Eqs. (3.25) through (3.28).

5. Calculate the inverse of interaction matrix $\bar{\bar{\mathbf{A}}}^{-1}$.

6. Calculate the correlation matrix of the incident field vector $\bar{\bar{\mathbf{R}}}_{\mathbf{EE}}$ using Eqs. (3.37) and (3.41).

7. Calculate the correlation matrix of the dipole moment vector $\bar{\bar{\mathbf{R}}}_{\mathbf{PP}}$ using Eq. (3.36).

8. Calculate the power absorbed $\langle Q_{abs,\omega} \rangle$ using Eq. (3.35).

## 4. Results and discussion

*4.1. Accuracy of the T-DDA*

The accuracy of the T-DDA is subjected to the same validity criteria as the DDA. According to Draine [62], there are three validity criteria associated with the DDA. The first condition stipulates that the number of sub-volumes should be large enough in order to describe the geometries of objects accurately [62]. The error introduced by this effect is known as the shape error. A general quantitative criterion providing the minimum number of sub-volumes to maintain the shape error within an acceptable range has not been established. Yurkin et al. [74] showed that shape error for a specific sub-volume size is bounded by a summation of $a_i$ and $a_i^2$ terms ($a_i$ is the effective dipole radius associated with a cubical sub-volume), but no information has been provided regarding the coefficients of these error terms. Draine [62] suggested a minimum number of sub-volumes for a sphere in the zero-frequency limit:

$$N_{min} \approx 60|n-1|^3 \left(\frac{\Delta}{0.1}\right)^{-3} \quad (4.1)$$

where $n = \sqrt{\varepsilon}$ is the complex refractive index of the material while $\Delta$ is the desired fractional accuracy. Equation (4.1) shows that the minimum number of sub-volumes increases as the refractive index increases for a given $\Delta$ value.

According to the second criterion, the discretization should be small enough when compared to the wavelength in the material ($2\pi/n'k_0$) and when compared to the attenuation length of the wave inside the material ($2\pi/n''k_0$) [62]. The latter becomes more important for highly absorptive materials such as metals. A simple criterion for satisfying this condition has been suggested [62,63,75]:

$$N_{min} \approx \frac{4\pi}{3}|n|^3 (k_0 a_{eff})^3 \left(\frac{\Delta}{0.1}\right)^{-3} \quad (4.2)$$

where $a_{eff}$ is the effective radius of the object defined as $a_{eff} = (3V/4\pi)^{1/3}$.

The last condition is concerned with the fact that magnetic dipole effects are neglected when discretizing objects into sub-volumes behaving as electric point dipoles. Indeed, even for non-magnetic materials, magnetic dipole absorption might be comparable, or even greater, than electric dipole absorption when dealing with conductive media such as metals [62,76]. The relative importance of magnetic dipole effects reduces with decreasing the size of the sub-volumes [62]. Therefore, the T-DDA can be applied to conductive media provided that the discretization is fine enough. Draine [62] suggested a criterion that combines Eq. (4.2) with $C_{abs}^m / C_{abs}^e < \Delta$, where $C_{abs}^m$ and $C_{abs}^e$ are the magnetic dipole absorption cross-section and the electric dipole absorption cross-section, respectively. This criterion is given by:

$$N_{min} \approx \frac{4\pi}{3} |n|^3 (k_0 a_{eff})^3 \left(\frac{\Delta}{0.1}\right)^{-3} \left[1 + \frac{|n|^3}{36\pi}\left(\frac{\Delta}{0.1}\right)^{3/2}\right] \tag{4.3}$$

By comparing Eq. (4.3) against Eq. (4.2), it is clear that magnetic dipole effects should be considered in the discretization if the following condition is satisfied:

$$|n| \geq (36\pi)^{1/3} \left(\frac{\Delta}{0.1}\right)^{-1/2} \tag{4.4}$$

The T-DDA is evaluated next for different material properties, separation gaps and object shapes.

*4.2. Verification of the T-DDA*

The T-DDA is first tested by computing the spectral radiative conductance between two spheres and by comparing the results against those obtained from the analytical solution of Narayanaswamy and Chen [33]. The thermal conductance between two objects is defined as follows. Assuming that the first object is at temperature $T$ while the second object is at temperature $T + \delta T$, the spectral conductance $G_\omega$ is calculated as:

$$G_\omega = \lim_{\delta T \to 0} \frac{\langle Q_{net,\omega} \rangle}{\delta T} \tag{4.5}$$

where $Q_{net,\omega}$ is the net spectral heat transfer rate between the two objects. The case of two spheres of diameter $D = 200$ nm separated by a distance $d = 10$ nm is analyzed in Fig. 2; the temperature is fixed at 400 K in all cases considered. The spheres of Fig. 2(a) have a frequency-independent dielectric function of $\varepsilon = 1.2 + 0.1i$, while the dielectric functions of the spheres in Figs. 2(b) and 2(c) are assumed to be $\varepsilon = 2.5 + 0.1i$ and $\varepsilon = 7.0 + 0.1i$, respectively. The minimum number of sub-volumes, $N_{min}$, required for satisfying the criteria discussed in section 4.1 are shown in Table 1 for an accuracy $\Delta = 0.01$. It can be seen in Table 1 that the criterion given by Eq. (4.1) dictates the minimum number of sub-volumes required. Furthermore, the condition given by Eq. (4.3) is ignored since Eq. (4.4) is never satisfied for all cases treated in this paper. The T-DDA simulations have been performed for five different numbers of sub-volumes, from 8 to 552 in each sphere (shown in parentheses in the figures).

[Insert Table 1 about here]

Figure 2(a) shows that for spheres with dielectric functions close to unity, the T-DDA approaches the analytical solution as the number of sub-volumes increases, and eventually converges to the exact results for 280 sub-volumes. This number is greater than the predicted value given in Table 1. A different behavior is observed in Fig. 2(b) for a larger dielectric function. As the number of sub-volumes increases, the T-DDA oscillates around the analytical solution. Figure 2(b) shows that the best results are obtained for 32 sub-volumes. This behavior is in agreement with the observations reported by Yurkin et al. [74], where results showed that the error does not monotonically decreases as the sub-volume size decreases. Instead, the error as a function of the spatial discretization displays a local minimum. For the largest dielectric function, an oscillatory behavior can again be observed in Fig. 2(c). The amplitude of these oscillations is larger than in the previous case of Fig. 2(b). There is a nearly perfect match between the T-DDA and the analytical solution when 136 sub-volumes are used. The comments made for Fig. 2(b) are therefore applicable to the case of Fig. 2(c).

[Insert Fig. 2 about here]

The T-DDA is also applied for calculating the conductance between two spheres made of silica. The diameters of the spheres are fixed at 0.5 μm while the temperature is kept at 400 K. Figures 3(a) and 3(b) show conductance profiles for separation gaps of 0.5 μm and 0.2 μm, respectively. The frequency-dependent dielectric function of silica has been extracted from the data reported in Ref. [77]. According to the criteria given by Eqs. (4.1) and (4.2), a minimum number of 688160 and 1193 sub-volumes in each sphere is required to ensure an accuracy of $\Delta = 0.01$, respectively, regardless of the separation gap $d$. However, it can be seen in Figs. 3(a) and 3(b) that a good agreement between the T-DDA and the analytical solution is achieved for a significantly smaller number of sub-volumes. Additionally, the resonant frequencies are predicted accurately via the T-DDA. For 552 sub-volumes in each sphere, the relative differences between the total conductances calculated from the T-DDA and the analytical solution within the spectral band of from 0.04 eV to 0.16 eV are 0.35% and 6.4% for separation gaps of 0.5 μm and 0.2 μm, respectively. The criteria discussed in section 4.1 are therefore of little help for the cases treated here. Also, these criteria do not account for the separation distance between the objects, while the results of Figs. 3(a) and 3(b) suggest that for a fixed number of sub-volumes, the error increases as the separation gap decreases. Further investigation is needed for developing criteria more appropriate to near-field thermal radiation calculations; this is left as a future research effort as it is beyond the scope of this paper.

[Insert Fig. 3 about here]

Finally, the spectral conductance between two cubes of silica calculated with the T-DDA is shown in Fig. 4. Both cubes have side lengths of 0.5 μm and are separated by a distance of 0.5 μm; as before, the temperature is fixed at 400 K. Note that an analytical solution does not exist for this problem. Additionally, Eq. (4.1) cannot be applied to cubes for estimating the minimum number of sub-volumes required to ensure an accuracy of $\Delta = 0.01$. Nevertheless, a fast convergence of the T-DDA is observed in Fig. 4. Indeed, the relative difference between the total conductances for 64 and 216 sub-volumes in each cube is 1.5%, while the relative difference is 0.6% between 216 and 512 sub-volumes (result not shown). This is in agreement with Yurkin et al. [74], where it was shown that the error associated with cubically shaped objects is smaller than for curved geometries.

[Insert Fig. 4 about here]

## 5. Conclusions

A new method called the Thermal Discrete Dipole Approximation (T-DDA) was proposed for modeling near-field thermal radiation in three-dimensional arbitrary geometries. The T-DDA is conceptually similar to the Discrete Dipole Approximation, except that the incident field originates from thermal oscillations of dipoles rather than by an external radiation source. The T-DDA was verified against exact results for two spheres separated by a sub-wavelength gap for various dielectric functions and separation distances. In all cases considered, the results obtained from the T-DDA were in good agreement with the exact results. Moreover, the resonant frequencies arising between two silica spheres were predicted accurately. For this last case, the relative errors between the analytical and the T-DDA results of total conductance were 0.35% and 6.4% for two 0.5 μm-diameter spheres separated by gaps of 0.5 μm and 0.2 μm, respectively. The T-DDA results exhibited an oscillatory behavior when the dielectric function of the materials is far from unity. It was also observed that for a fixed number of sub-volumes, the error increases as the separation gap decreases. Additionally, faster convergence was achieved when considering two cubes rather than two spheres.

This paper suggests that the T-DDA is a robust, relatively simple simulation tool for predicting near-field radiative heat exchange. The convergence and the accuracy of the T-DDA need further investigation, and this is left as a future research effort.

**Acknowledgment**

This work was partially supported by the US Army Research Office under Grant No. W911NF-12-1-0422 Mod. 1.

[26] Narayanaswamy A. and Chen G., Direct computation of thermal emission from nanostructures, *Annual Reviews of Heat Transfer* **14**, 169-195, 2005.

[27] Francoeur M., Mengüç M.P. and Vaillon R., Solution of near-field thermal radiation in one-dimensional layered media using dyadic Green's functions and the scattering matrix method, *Journal of Quantitative Spectroscopy and Radiative Transfer* **110**, 2002-2018, 2009.

[28] Zheng Z. and Xuan Y., Theory of near-field radiative heat transfer for stratified magnetic media, *International Journal of Heat and Mass Transfer* **54**, 1101-1110, 2011.

[29] Hammonds Jr. J.S., Thermal transport via surface phonon polaritons across a two-dimensional pore, *Applied Physics Letters* **88**, 041912, 2006.

[30] Domingues G., Volz S., Joulain K. and Greffet J.-J., Heat transfer between two nanoparticles through near field interaction, *Physical Review Letters* **94**, 085901, 2005.

[31] Chapuis P.-O., Laroche M., Volz S. and Greffet J.-J., Radiative heat transfer between metallic nanoparticles, *Applied Physics Letters* **92**, 201906, 2008.

[32] Chapuis P.-O., Laroche M., Volz S. and Greffet J.-J., Erratum: "Radiative heat transfer between metallic nanoparticles", *Applied Physics Letters* **97**, 269903, 2010.

[33] Narayanaswamy A. and Chen G., Thermal near-field radiative transfer between two spheres, *Physical Review B* **77**, 075125, 2008.

[34] Krüger M., Bimonte G., Emig T. and Kardar M., Trace formulas for nonequilibrium Casimir interactions, heat radiation, and heat transfer for arbitrary objects, *Physical Review B* **86**, 115423, 2012.
28

**Figure captions:**

Figure 1. Schematic representation of the problem under consideration.

Figure 2. Verification of the T-DDA against analytical results for two spheres with dielectric functions of: (a) $\varepsilon = 1.2 + 0.1i$, (b) $\varepsilon = 2.5 + 0.1i$, and (c) $\varepsilon = 7.0 + 0.1i$. In all panels, the temperature $T$ is fixed at 400 K.

Figure 3. Spectral conductance between two silica spheres for separation gaps of: (a) 0.5 μm, and (b) 0.2 μm. In both panels, the T-DDA is compared against analytical results and the temperature $T$ is fixed at 400 K.

Figure 4. Spectral conductance between two silica cubes separated by a 0.5 μm-thick gap. The temperature $T$ is fixed at 400 K.

**Table caption:**

Table 1. Minimum number of sub-volumes required for satisfying the criteria given by Eqs. (4.1) and (4.2).

|  | $N_{min}$ from Eq. (4.1) | $N_{min}$ from Eq. (4.2) |
|---|---|---|
| Fig. 2(a) | 72 | 46 |
| Fig. 2(b) | 11850 | 138 |
| Fig. 2(c) | 267530 | 645 |
| Figs. 3(a) and 3(b) | 688160 | 1193 |
| Fig. 4 | N/A | 2279 |

Figure 1

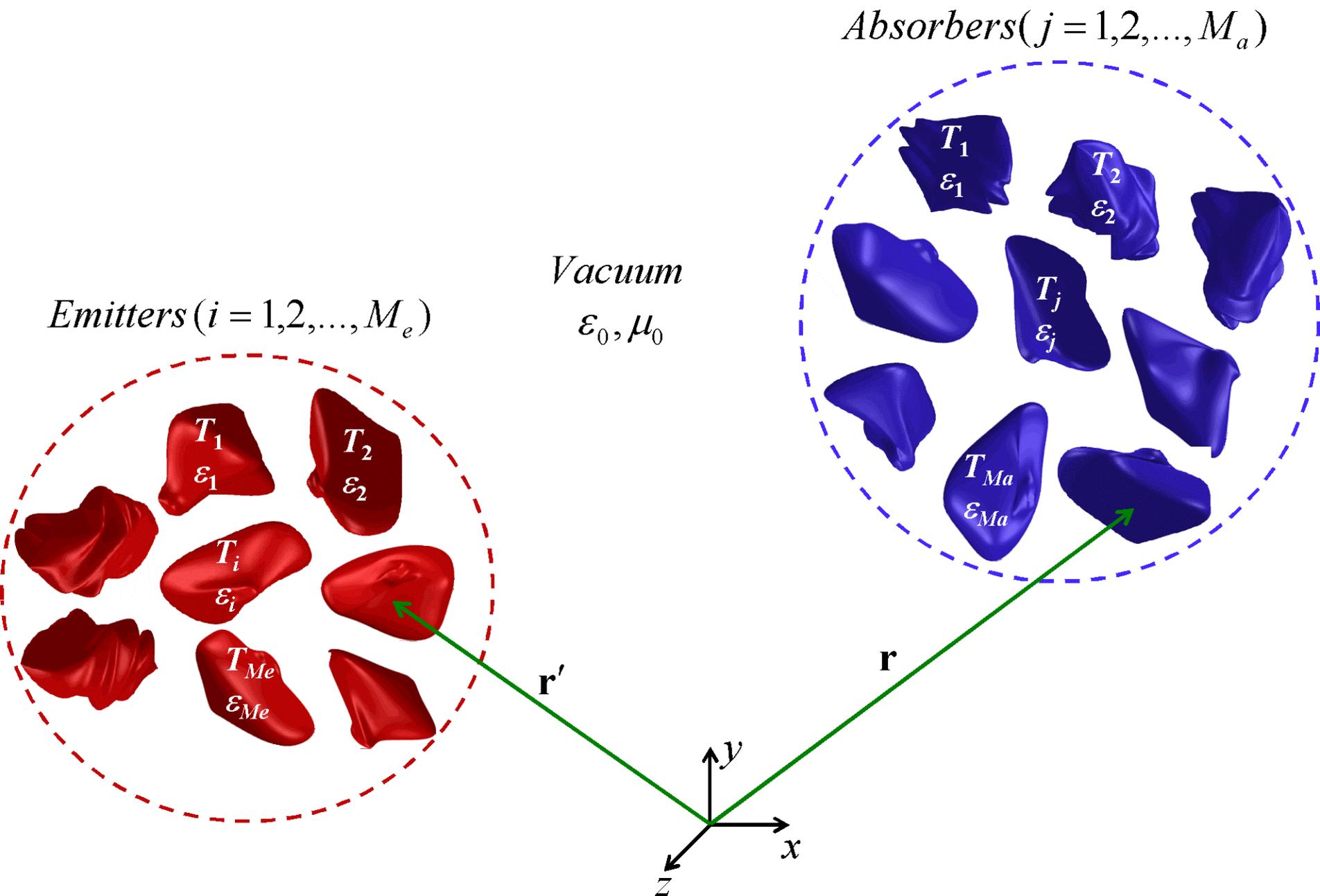

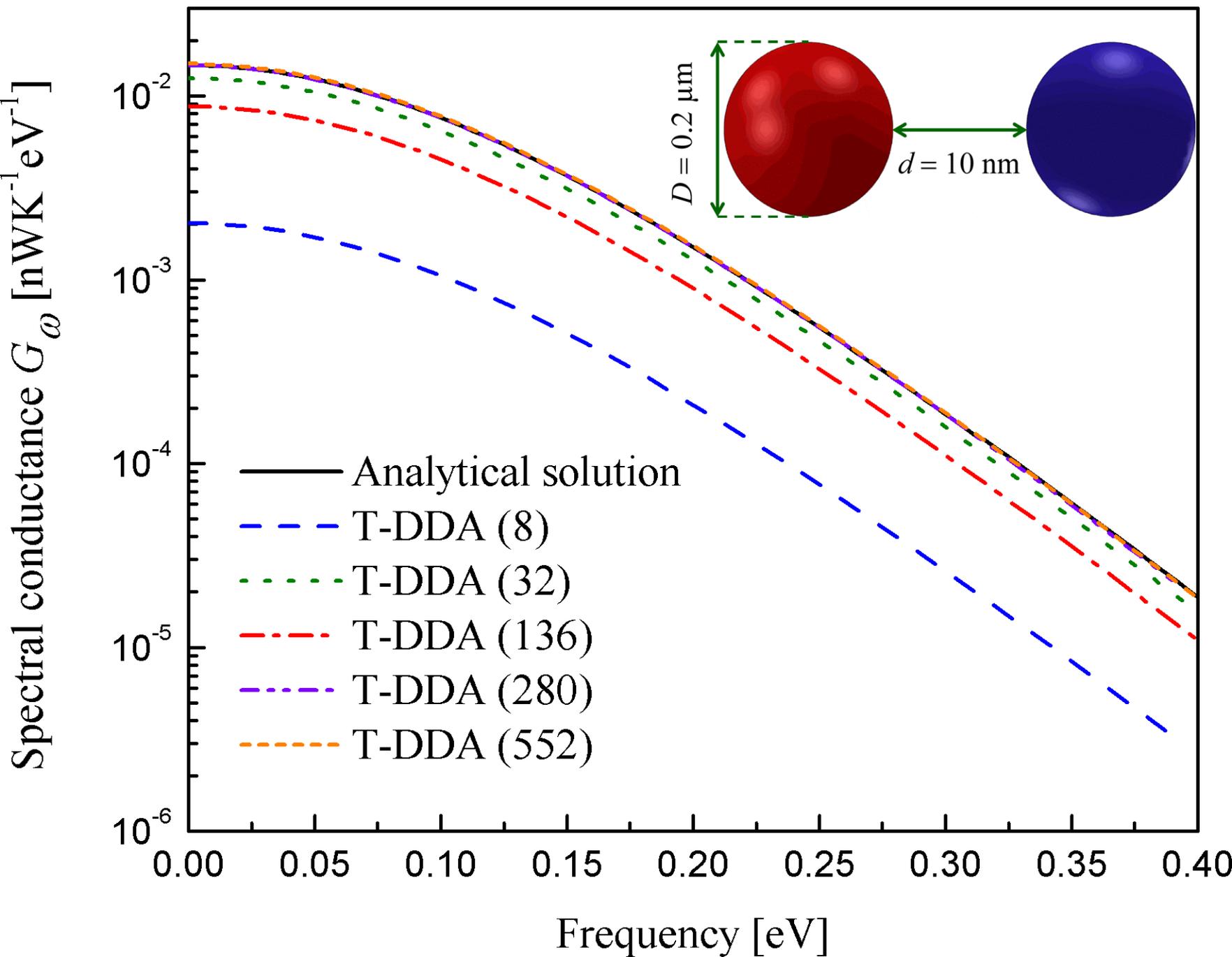

Figure 2a

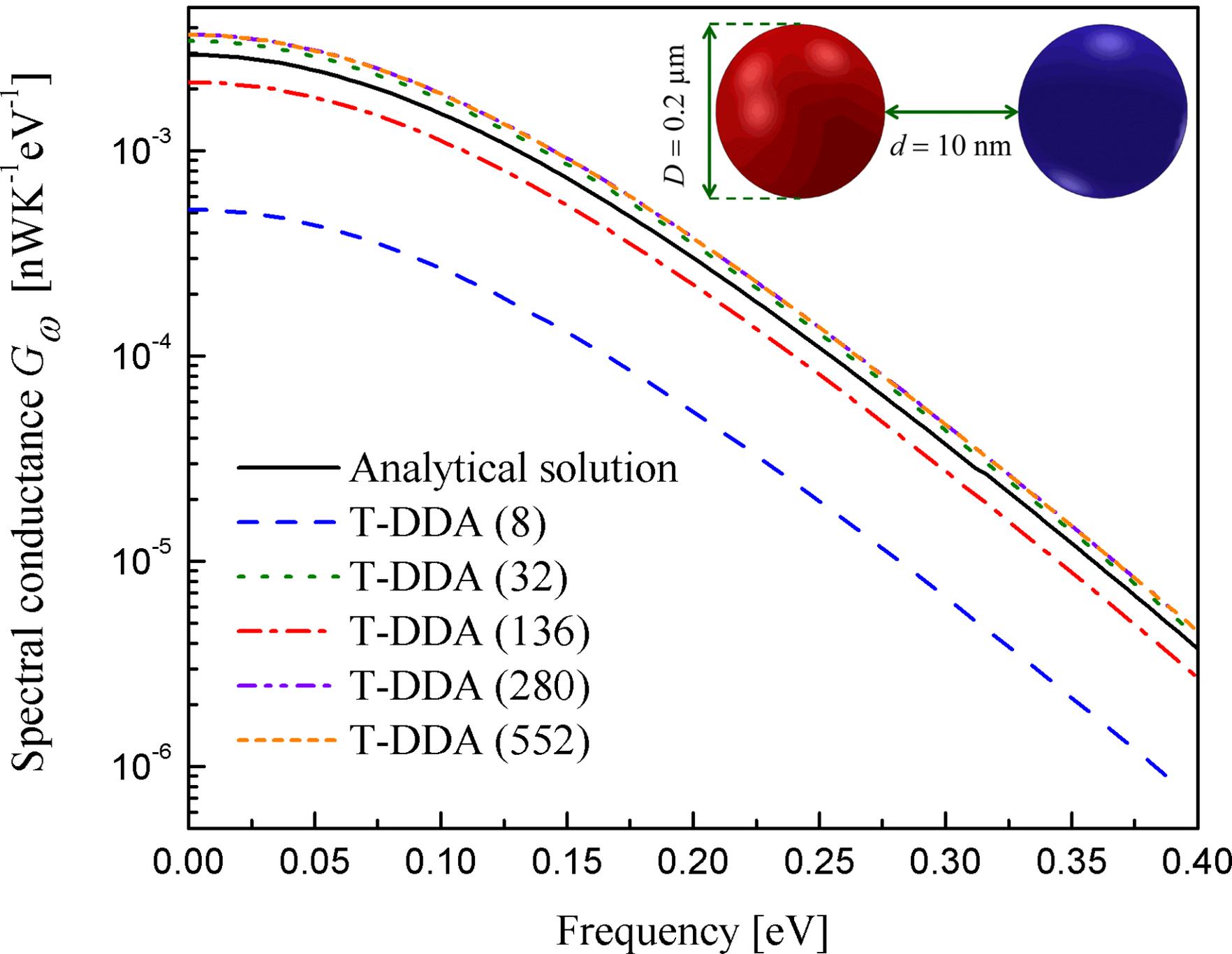

Figure 2b

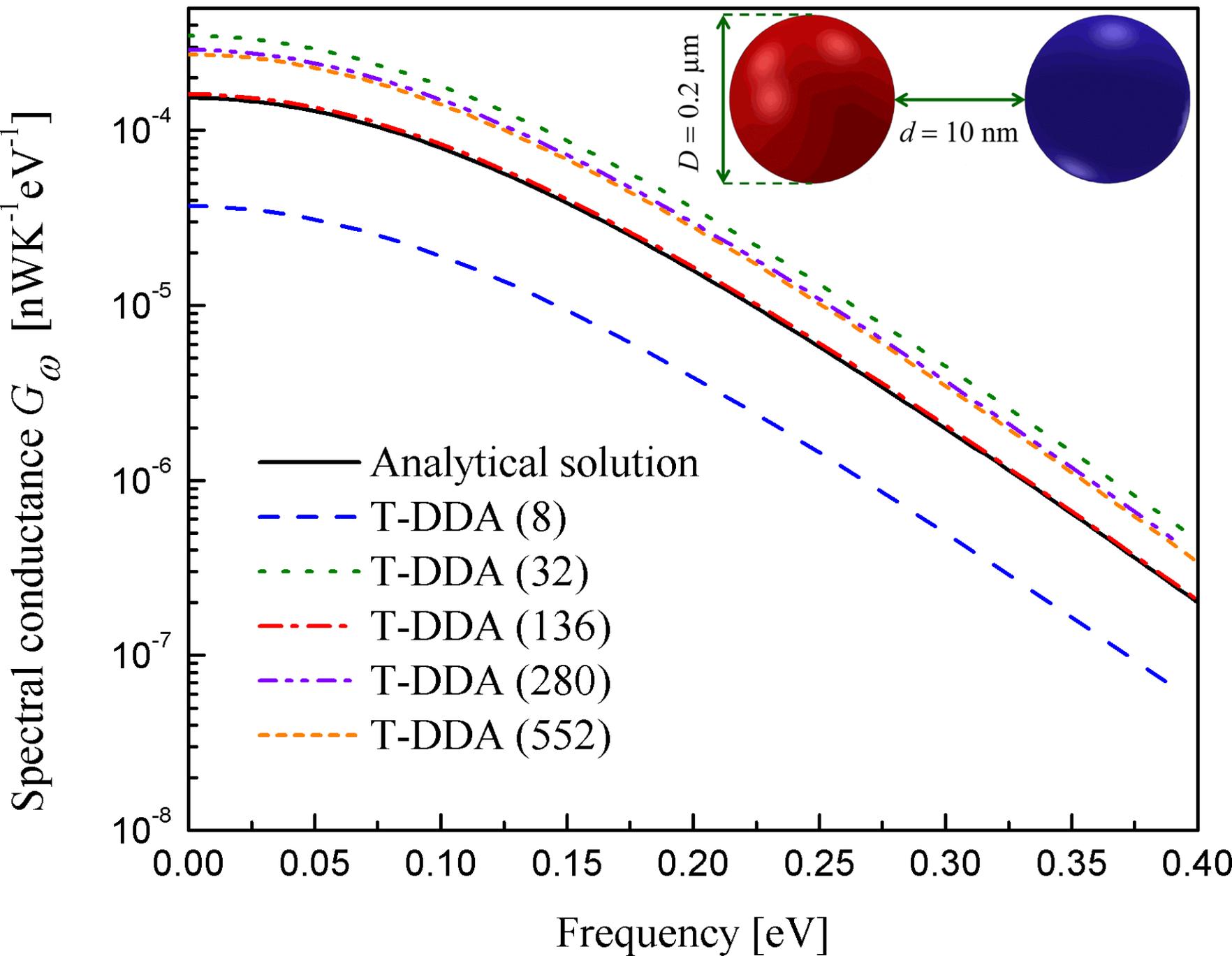

Figure 2c

Figure 3a

*Spectral conductance $G_\omega$ [nWK$^{-1}$eV$^{-1}$] vs Frequency [eV]. Legend: Analytical Solution; T-DDA (8); T-DDA (32); T-DDA (136); T-DDA (280); T-DDA (552). Inset: two spheres with $D = 0.5\ \mu m$ and $d = 0.5\ \mu m$.*

Figure 3b

Spectral conductance $G_\omega$ [nWK$^{-1}$eV$^{-1}$] vs Frequency [eV]

Legend:
- Analytical Solution
- T-DDA (8)
- T-DDA (32)
- T-DDA (136)
- T-DDA (280)
- T-DDA (552)

$D = 0.5\ \mu m$, $d = 0.2\ \mu m$

Figure 4

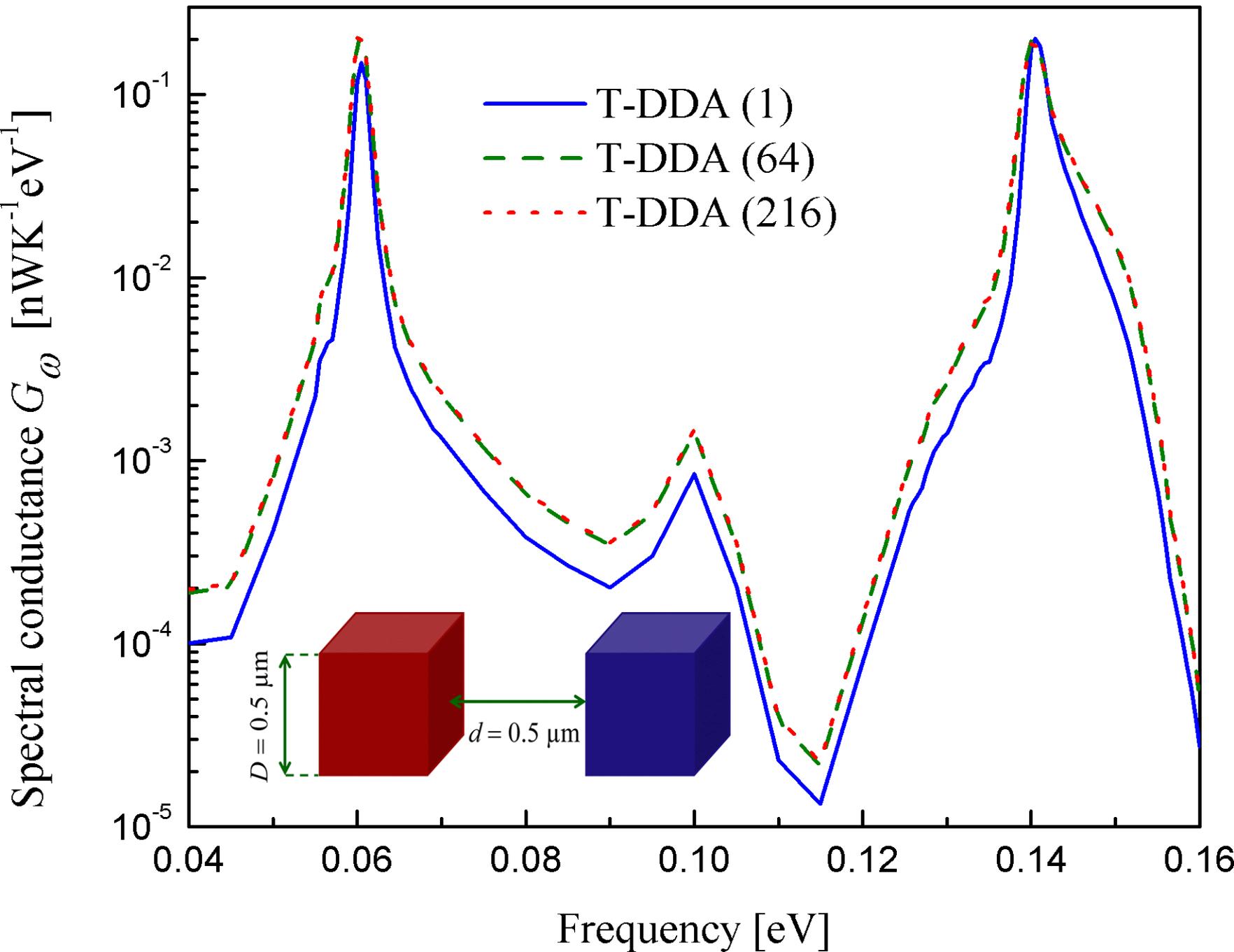